\documentclass{aastex}

\usepackage{graphicx}
\usepackage{amsmath}
\usepackage{amssymb}
\usepackage{natbib}
\usepackage{xspace}
\usepackage{relsize}

\newcommand{\ten}[1] {10\ensuremath{^{#1}}\xspace}

\shorttitle{Rotational Excitation Temperature of $\lambda$6614 DIB}
\shortauthors{Cami et al.}

\begin{document}
\title{The Rotational Excitation Temperature of the $\lambda$6614
  Diffuse Interstellar Band Carrier}

\author{J.~Cami\altaffilmark{1}, F.~Salama\altaffilmark{1}, 
  J.~Jim\'enez-Vicente\altaffilmark{2}, 
  G.A.~~Galazutdinov\altaffilmark{3,4} and J.~Krelowski\altaffilmark{5}}
\altaffiltext{1}{NASA Ames Research Center, MS 245-6, Moffett Field,
  CA 94035, USA}\email{jcami@mail.arc.nasa.gov}
 \email{fsalama@mail.arc.nasa.gov}
\altaffiltext{2}{Depto.~F\'\i sica Te\'orica y del Cosmos, Fac.~de
  Ciencias, Univ.~de Granada, Av.~Fuentenueva s/n, 18071 Granada, Spain}
  \email{jjimenez@ugr.es}
\altaffiltext{3}{Korea Astronomy Observatory, Optical Astronomy Division,
61-1, Whaam-Dong, Yuseong-Gu, Daejon 305-348, Korea}
  \email{gala@boao.re.kr}
\altaffiltext{4}{Special Astrophysical Observatory, Nizhnij Arkhyz 369167,
Russia}
\altaffiltext{5}{Center for Astronomy, Nicolaus Copernicus University, Gagarina
  11, 87-100 Tor\'un, Poland}
\email{Jacek.Krelowski@astri.uni.torun.pl}

\begin{abstract}
 Analysis of high spectral resolution observations of the
 $\lambda$6614 diffuse interstellar band (DIB) line profile show
 systematic variations in the positions of the peaks in the
 substructure of the profile. These variations -- shown here for the
 first time -- can be understood most naturally in the framework of
 rotational contours of large molecules, where the variations are
 caused by changes in the rotational excitation temperature. We show
 that the rotational excitation temperature for the DIB carrier is
 likely significantly lower than the gas kinetic temperature --
 indicating that for this particular DIB carrier angular momentum
 buildup is not very efficient.
\end{abstract}


\section{Introduction}

The diffuse interstellar bands (DIBs) are over 300 interstellar
absorption bands commonly observed toward reddened stars from the UV
to the near-IR, and whose carrier molecules are still unidentified
\citep{Herbig:1995,Krelowski:reviewErice2000}. The identification of
the carriers of these bands remains an important problem in astronomy
to date and the current consensus on the nature of the carriers is
that they are probably large carbon-bearing molecules that reside
ubiquitously in the interstellar gas
\citep{Ehrenfreund-charnley-review}. The most promising carrier
candidates are carbon chains, polycyclic aromatic hydrocarbons (PAHs),
and fullerenes \citep{Salama:1996,Salama:1999,Foing:C60_1,%
Schulz:chains:contours,Motylewski:chains}. Studies on the
environmental behavior of DIB carriers suggest that the strength of
the DIBs results from an interplay between ionization, recombination,
dehydrogenation and destruction of chemically stable, carbonaceous
species \citep{Cami:DIBcorrelations,%
Sonnentrucker:ionization,Vuong-Foing}. The molecular nature of the DIB
carriers is supported by the detection of substructures in the line
profiles of some DIBs \citep{Sarre:High_resolution,%
Ehrenfreund:substructures,Krelowski-Schmidt:profiles,%
Walker:profiles}.
Furthermore, a recent analysis of the profile of the strongest DIB
($\lambda$4428) shows a Lorentzian profile remarkably consistent with
rapid internal conversion in a molecular carrier
\citep{Snow:4430profile}.

In this Letter, we present an analysis of high-resolution observations
of the $\lambda$6614 DIB in so-called single-cloud lines of
sight. The profiles show a systematic variation in the wavelengths of
the observed substructure peaks. We show that these variations are
most naturally explained as changes in the rotational excitation
temperature in rotational contours of a large molecule. The particular
band profile of this DIB also allows one to uniquely determine the
rotational excitation temperature.

\section{The $\lambda$6614 DIB profile}

The $\lambda$6614 DIB was first observed at high resolution by
\citet{Sarre:High_resolution}, revealing a clear triple-peak
substructure and a red degraded wing (see
Fig.~\ref{Fig:DIB6614:allprofiles}). In a few stars, a weaker fourth
and a fifth peak show up at longer wavelengths. As these are only
clearly observed in two of our spectra, we will not discuss these
peaks in this Letter. The substructures are intrinsic to the band
profile, as the same profile shape is observed in lines of sight that
only cross one interstellar cloud.

The very presence of these substructures has been explained by two
different scenarios. The most popular explanation is that the profile
is due to unresolved rotational contours of a large molecule, in which
the three peaks represent individual branches of a rovibronic
transition. Rotational contour calculations have been performed by
\citet{Cossart-Magos:Contours} and \citet{Kerr:6613DIB} for PAHs,
by \citet{Edwards-Leach:C60:countours} for fullerenes and by
\citet{Schulz:chains:contours} for linear
chains. \citet{Ehrenfreund:substructures} analyzed the profiles of the
$\lambda$6614 DIB and compared them to calculated rotational contours,
concluding that the carrier of the $\lambda$6614 DIB has a rotational
constant compatible with PAHs larger than 40 C atoms, chains of 12-18
C atoms, 30 C atom rings, or C$_{60}$ fullerene compounds.

Alternatively, the substructures might be due to isotope shifts in
large, highly symmetric molecules \citep{Webster:isotopeshift}. In
this scenario, the individual peaks correspond to entities of the same
molecule with a different number of $^{13}$C atoms. The relative
intensities of the peaks then determine the abundances of the isotopic
varieties \citep[see, e.g.,][]{Walker:6614:isotope}. 

\section{The $\lambda$6614 DIB profile variations}

\begin{figure}
\begin{center}
\resizebox{6.3cm}{!}{\includegraphics{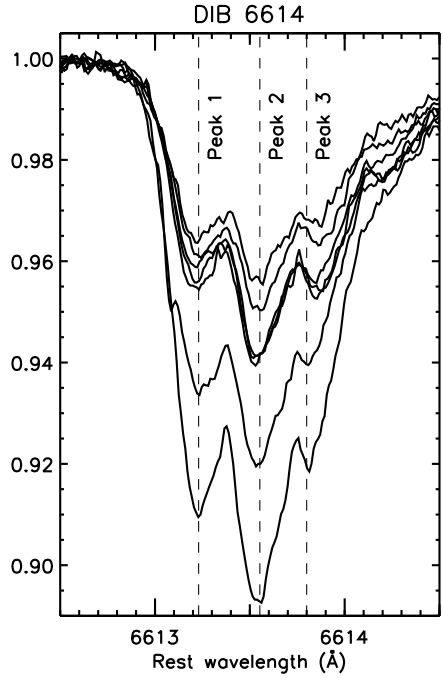}}
\caption{\label{Fig:DIB6614:allprofiles}Profiles of the
  $\lambda$6614 DIB for the observed stars. All profiles are centered
  on the central peak (peak 2). Note how the position of the redward
  peak (peak 3) clearly changes from one line of sight to another. At
  the same time, the width of the sub peaks seems to become marginally
  larger. For an identification of the individual spectra, see
  \citet{Galazutdinov:LP6614:LP6196}.}
\end{center}
\end{figure}

Recently, \citet{Galazutdinov:LP6614:LP6196} presented new
high-resolution ($R \approx$ 220000), high signal-to-noise ratio
observations of the $\lambda$6614 DIB toward single-cloud
lines of sight, clearly showing variability in the precise wavelengths
of the peaks, and in the intensity ratios. In this Letter, we analyze
those same observations; we therefore refer to
\citet{Galazutdinov:LP6614:LP6196} for more observational details. 

The profiles of the interstellar \ion{Na}{1} or \ion{K}{1} lines
toward our target stars are narrow, and generally show only one main
component \citep{Galazutdinov:LP6614:LP6196}, confirmed also by
ultra-high resolution observations for various interstellar lines
that are now available \citep[see e.g.][and references
therein]{Welty:CaI-Hires}. The observed DIB profiles are therefore
intrinsic.

Fig.~\ref{Fig:DIB6614:allprofiles} shows the profiles of the
$\lambda$6614 DIB centered on the central peak. It is clear that there
are considerable variations in the exact positions of the remaining
two main peaks with respect to this central peak. The redward peak
(peak 3) shows clear variations in wavelength relative to the central
peak that are much larger than the individual variations in radial
velocity, and therefore these variations are intrinsic. To assess the
nature of these variations, we proceeded in the following way.  In all
proposed explanations for the substructure of the $\lambda$6614 DIB,
the observed profile is composed of individual bands. In the
rotational contour framework, these bands correspond to unresolved
ro-vibronic branches; in the isotope shift scenario they correspond to
different isotopomers. To accurately determine the peak positions of
these individual components, we need to first decompose the profile
into its components, rather than measuring the peak positions directly
in the observed profile.

\begin{figure}[t!]
\begin{center}
\resizebox{\hsize}{!}{\includegraphics{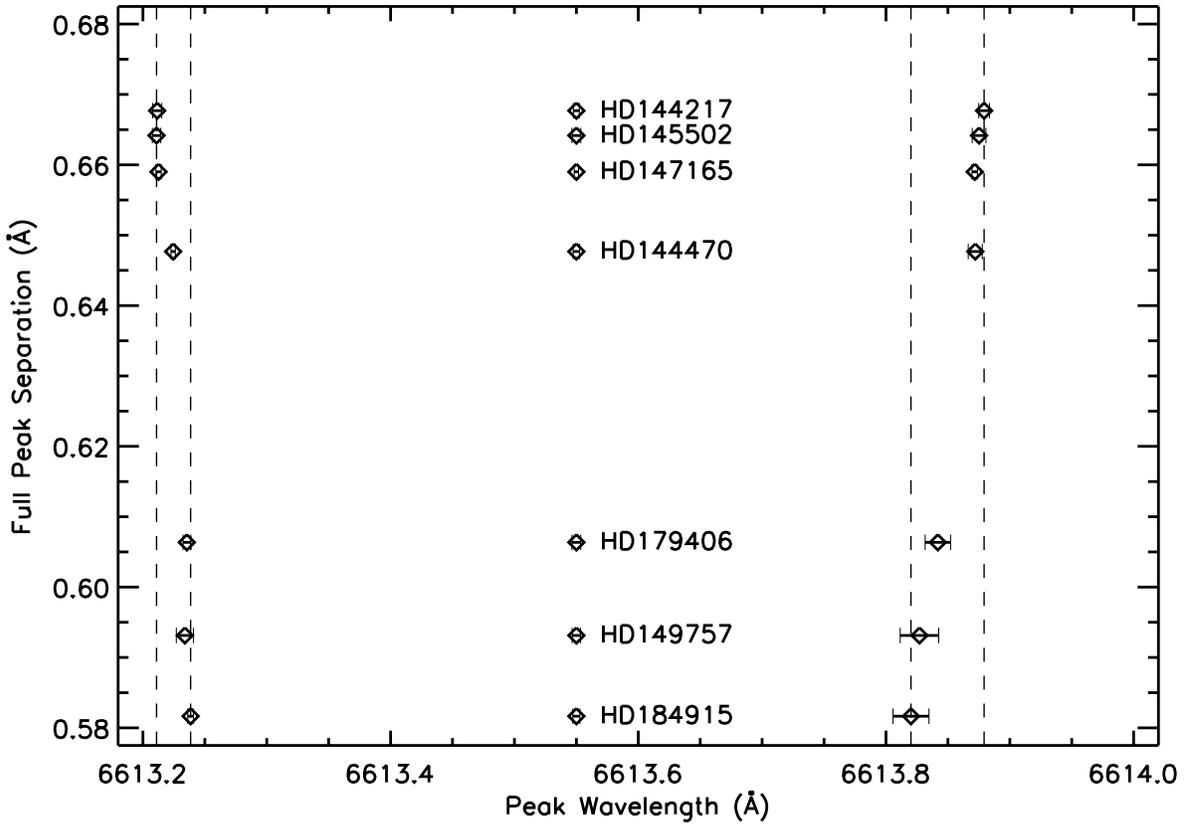}}
\caption{\label{Fig:DIB6614:peakshifsts}Peak positions for the
  three main peaks in the $\lambda$6614 DIB profile. Error bars are
  determined using a statistical approach (see text).}
\end{center}
\end{figure}

\citet{Galazutdinov:LP6614:LP6196} showed that nearly perfect fits to
the observed line profiles can be achieved by fitting about five
Gaussians to the line profiles. The three main Gaussian components
correspond to the three main peaks observed in the line profile; a
fourth and fifth component are required for the additional peaks
and/or the red wing. We followed a similar approach to decompose the
profile into Gaussian components. The fitting routine provides a
formal error estimate on the derived parameters (such as peak
position) which is generally optimistic since it does not deal with
systematic errors -- such as the components not being true
Gaussians. We therefore followed a statistical approach in which we
determined the noise level on the spectra from the continuum parts
outside the $\lambda$6614 DIB, subsequently added random noise with
the same rms to the entire spectrum, and then fitted the
profile. This procedure was repeated 100 times for each spectrum, and
subsequently each parameter value was taken to be the average of these
100 simulations, with the standard deviation providing a more
realistic error estimate on the parameters. Using the wavelengths of
the central peak (peak 2 in Fig.~\ref{Fig:DIB6614:allprofiles}), we
then shifted the individual profiles in velocity space to the same
``rest'' wavelength of 6613.56 \AA\ for the central peak
\citep{Galazutdinov:DIBwavelengths}.

Fig.~\ref{Fig:DIB6614:peakshifsts} shows the peak positions for each
of the three main peaks as a function of the peak separation between
the first and the third peak.  The variations in the wavelengths of
the sub-peaks now become much more obvious. When the redward peak
(peak 3) shifts to longer wavelengths relative to the central peak,
the blueward peak (peak 1) {\em systematically} shifts to shorter
wavelengths. Whereas peak 1 shifts by about 0.03 \AA (one resolution
element), peak 3 shifts by twice that amount. 

These variations rule out the scenario in which the substructures are
due to isotope shifts; in such a case, the wavelengths of the
individual peaks should be the same for each line of sight.  On the
other hand, the observed variations are completely consistent with the
substructure being due to unresolved rotational contours and the
observed variations due to changes in only one \mbox{parameter :} the
rotational excitation temperature.

\section{Rotational contours}
\label{Sec:Rotational_contours}

In the framework of rotational contours, the three peaks in the
$\lambda$6614 DIB correspond to unresolved $PQR$-type branches
associated with a molecular species. For any molecular species, the
peak positions of these branches are determined by both molecular
properties and physical parameters (such as the rotational
temperature). The added value from this work -- where we observe {\em
variations} in the peak positions -- stems from the fact that the
molecular properties should be the same for all lines of
sight. Moreover, the triple-peak structure of the $\lambda$6614 DIB
profile allows one to determine the molecular and physical parameters
independently. This is most easily illustrated for molecules
exhibiting a linear or spherical top geometry.
%
%

For linear (e.g., ~CO$_2$, N$_2$O) or spherical top (e.g., fullerenes)
geometries, the energies of the rotational levels within a vibronic
band are determined by the single rotational constant $B$ and the
quantum number $J$ (angular momentum). Selection rules on $J$ are
$\Delta J=\pm 1$ ($P$- and $R$-branch transitions) and, in some cases
(see Sect.~\ref{Sec:discussion}), $\Delta J=0$ ($Q$-branch). For a
given rotational level $J$, the frequencies of the $P$ and $R$
transitions relative to the corresponding $Q$ transition can then be
written as
\begin{eqnarray}
\label{Eq:nu_RQ}
\Delta \nu_{RQ} = & 2(J+1)(B'' + \Delta B) & \approx 2(J+1)B'' \\
\label{Eq:nu_QP}
\Delta \nu_{QP} = & 2J(B'' + \Delta B) & \approx 2JB''
\end{eqnarray}
where $\Delta\nu_{RQ} \equiv \nu_R - \nu_Q$ and $\Delta\nu_{QP} \equiv
\nu_Q - \nu_P$; the double primes refer to the lower vibronic level
and $\Delta B$ is the difference in rotational constants between the
upper and lower vibronic level. Furthermore, the common approximation
has been used that $\Delta B/B'' << 1$.  All other things being equal,
the peak absorption will arise from the most populated rotational
level in the lower vibronic state.  Assuming an LTE-like population
distribution of the rotational levels in this lower state, it is
straightforward to show that the most populated rotational level is
\citep[see e.g.][]{Ehrenfreund:substructures}
\begin{equation}
\label{Eq:Jmax}
J_{\rm max} = \sqrt{\frac{kT_{\rm rot}}{2hcB''}} - \frac{1}{2}
\end{equation}
with $B''$ in units of cm$^{-1}$.  Eqs.~(\ref{Eq:nu_RQ}),
(\ref{Eq:nu_QP}) and (\ref{Eq:Jmax}) nicely show how the peaks of the
$R$- and $P$- branches move away from the $Q$-branch peak (at a
different rate) when $T_{\rm rot}$ increases; as $T_{\rm rot}$
increases, $J_{\rm max}$ becomes larger, and therefore both peak
separations increase. This corresponds exactly to what we observe (see
Fig.~\ref{Fig:DIB6614:peakshifsts}), and therefore the central peak
(peak 2) in the $\lambda$6614 DIB must correspond to the unresolved
$Q$-branch and peaks 1 and 3 to the $R$- and $P$-branches,
respectively.

The triple-peak structure of the $\lambda$6614 DIB profile offers 14
independent measurements ($\Delta\nu_{RQ}$ and $\Delta\nu_{QP}$ for
seven lines of sight) for eight free parameters ($B''$ and seven
rotational temperatures), so that all parameters can be uniquely
determined.  Although it is possible to determine $B''$ and $T_{\rm
rot}$ independently for each line of sight by manipulating
Eqs.~(\ref{Eq:nu_RQ})--(\ref{Eq:Jmax}), the uncertainties on the
derived parameters are large. Instead, we performed a $\chi^2$
minimization to determine the eight parameters that provide the best fit
to the 14 observed peak separations and determined the 1 $\sigma$
uncertainties on these parameters. The best-fit parameters are listed
in Table~\ref{Table:Peakpositions2} and yield a reduced $\chi^2$-value
of 1.45, indicating a good, but not a perfect fit. The derived
rotational constant (0.016 $\pm$ 0.003 cm$^{-1}$) is compatible with
published values for linear and aromatic molecules, e.g., C$_9$
\citep[0.014 cm$^{-1}$; ][]{1993JChPh..98.6678V}, HC$_7$N$^+$
\citep[0.018 cm$^{-1}$; ][]{2000JChPh.112.8899S}, C$_{13}$H$_{9}$N or
C$_{15}$H$_{9}$N \citep[0.018 cm$^{-1}$; ][]{Mattioda:PANHs}.

\renewcommand{\arraystretch}{1.5}
\begin{deluxetable}{lrrrrr}
\tabletypesize{\scriptsize}
\tablecaption{\label{Table:Peakpositions2}The Observed Peak
  Separations in the $\lambda$6614 DIB and Derived $T_{\rm
  rot}$-Values}
\tablehead{ \colhead{Star} &
\colhead{$\Delta\nu_{12}$} & \colhead{$\Delta\nu_{23}$} &
\colhead{$\Delta\nu_{13}$} & \colhead{$T_{\rm rot}$} &
  \colhead{$T_{\rm H_2}$\tablenotemark{a}} \\
\colhead{} & \colhead{(cm$^{-1}$)} & \colhead{(cm$^{-1}$)} &
  \colhead{(cm$^{-1}$)} & \colhead{K} & \colhead{K} \\
} 
\startdata 
HD 144217 & 0.774 & 0.753 & 1.527 & 25.5$^{+6}_{-4}$ & 88  \\ 
HD 145502 & 0.775 & 0.743 & 1.518 & 25.3$^{+6}_{-4}$ & 90  \\ 
HD 147165 & 0.771 & 0.735 & 1.507 & 24.9$^{+6}_{-4}$ & 64  \\ 
HD 144470 & 0.744 & 0.736 & 1.481 & 23.6$^{+6}_{-4}$ & 73  \\ 
HD 179406 & 0.719 & 0.667 & 1.386 & 21.5$^{+5}_{-4}$ & --  \\ 
HD 149757 & 0.723 & 0.633 & 1.356 & 21.2$^{+5}_{-4}$ & 54  \\ 
HD 184915 & 0.712 & 0.618 & 1.330 & 21.0$^{+5}_{-3}$ & 69  \\ 
\enddata
\tablecomments{$B''=16.4 \pm 3.1$ \ten{-3} cm$^{-1}$. Quoted
  Uncertainties are 1$\sigma$.}
\tablenotetext{a}{From \citet{Savage:H-survey}. For HD 179406, $T_{\rm
  H_2}$ is unknown from observations.}
\end{deluxetable}

For the more general case of symmetric rotors (we will not discuss the
case of asymmetric rotors), the rotational energies will depend on all
three rotational constants $A$, $B$, and $C$ (where by convention
\mbox{$A\ge B\ge C$}) and on the quantum number $K$, the component of
the angular momentum $J$ parallel to the symmetry axis of the
molecule. The constant $K$ can have values $-J,-J+1,\dots,J-1, J$ so
that each $J$ level is now split up into $2J+1$ levels with different
$K$-values. For prolate geometries (where $B=C$), these levels differ
in energy by $(A-B)K^2$; for oblate cases (where $A=B$,
e.g.~C$_6$H$_6$) they differ by $-(A-C)K^2$. Additional selection
rules are $\Delta K=0$ (``parallel'' transitions) or $\Delta K=\pm 1$
(``perpendicular'' transitions). The observed profile will now be the
superposition of all $PQR$-branches arising from different $K$-values.


For parallel transitions from the $K=0$ level, the additional energy
term vanishes, and therefore Eqs.~(\ref{Eq:nu_RQ}) and (\ref{Eq:nu_QP})
still yield the correct peak separations. For parallel transitions
from $K\ne 0$, the frequencies of the transitions will change by
$(\Delta A-\Delta B)K^2$ or $(\Delta A-\Delta C)K^2$ compared to the
$K=0$ frequencies. These shifts are generally small, and therefore the
overall shape of the profile in terms of peak positions will not
change much. In such a case, the estimates of the rotational constant
$B$ and the rotational temperature $T_{\rm rot}$ derived from the
linear or spherical top case will still be good estimates.

For perpendicular bands, the situation is more complex. For each $K$,
there are now two sets of $PQR$-branches. Compared to the linear or
spherical top case, the transitions in the first set shift to the blue
and in the second to the red by an amount of typically $2(A-B)K$ for
prolate tops or $2(A-C)K$ for oblate tops. The superposition of all
these branches now leads to a broadening of the observed branches,
most noticeable in the $Q$-branch. Moreover, the $P$- and $R$-branches
for oblate tops will move away from the $Q$-branch by $\sim 2J_{\rm
max}(A-C)$; those for prolate tops move closer to the $Q$-branch by
$\sim 2J_{\rm max}(A-B)$. When using the formalism for linear or
spherical top geometries to determine $T_{\rm rot}$, we will therefore
generally overestimate $T_{\rm rot}$ for oblate geometries and
underestimate $T_{\rm rot}$ for prolate geometries.


\section{Discussion}
\label{Sec:discussion}

The observed variations in the peak positions of the substructures in
the $\lambda$6614 DIB can be explained at least qualitatively in the
rotational contour framework where the variations are due to changes
in the rotational excitation temperature. The formalism for linear or
spherical top geometries furthermore yields convenient expressions to
uniquely determine $B''$ and $T_{\rm rot}$ from the observed peak
separations. Crucial in this formalism is that the central peak in the
observed $\lambda$6614 DIB line profile is the $Q$-branch. However,
linear and spherical top molecules only exhibit $Q$-branches for very
specific vibronic transitions. In those cases, the $Q$-branch is
generally much narrower than the $P$- and $R$-branches and in many
cases also much stronger. The $\lambda$6614 DIB, on the other hand,
shows a central peak that has by and large the same width as the $P$-
and $R$-branches, and a comparable strength. It seems therefore
unlikely that the $\lambda$6614 DIB carrier conforms to a linear or
spherical top geometry. Rather, the broadening of the $Q$-branch is
presumably due to the superposition of the various stacks in a prolate
or oblate top molecule. As discussed in
Sect.~\ref{Sec:Rotational_contours}, this means the values for $T_{\rm
rot}$ in Table~\ref{Table:Peakpositions2} are either too low (prolate)
or too high (oblate). It is interesting to note in this context that
the detailed rotational contour calculations by \cite{Kerr:6613DIB}
for planar oblate tops (where $B=2C$) do indeed reproduce the width
and strength of the observed subpeaks in the $\lambda$6614 DIB
profile. Such detailed model calculations to compare to the
$\lambda$6614 DIB variations are in progress and will be presented in
a future paper. However, for the value of $B''$ we derived, the
rotational temperatures from \citet{Kerr:6613DIB} do indeed indicate
slightly lower values than those in Table~\ref{Table:Peakpositions2}.

It is therefore tempting to conclude that the rotational temperatures
for the $\lambda$6614 DIB carrier are indeed relatively low and, as
indicated in Table~\ref{Table:Peakpositions2}, significantly lower
than the kinetic gas temperature in the same lines of sight.  This is
somewhat surprising, as it has been argued that the rotational
excitation temperature should actually be higher than the gas
temperature
\citep[see e.g.][]{Rouan:PAHrotation,Malloci:profiles}. 
Clearly, either the rotational excitation processes (rocket effect,
intramolecular vibration-rotation energy transfer, ...) included in
those calculations are less important than assumed or the relaxation
processes are more efficient. However, these calculations are
generally carried out for large ($\sim$ 100 C atoms) PAH-like
molecules. Smaller molecules or molecules of a different geometry
might show a different excitation and relaxation balance. The
rotational constant we derive here is indeed indicative of smaller
molecules.

\section{Conclusions}

We have analyzed high-resolution and high signal-to-noise ratio
spectra of the $\lambda$6614 DIB toward single-cloud lines of
sight. The spectra clearly show a systematic shift in the relative
peak position of the subpeaks, shown here for the first time. This
cannot be understood if the substructure is due to isotope shifts. On
the other hand, this effect can be explained both qualitatively and
quantitatively by rotational contours in which only the rotational
excitation temperature changes.
The rotational excitation temperatures are likely to be lower than the
kinetic gas temperature, indicating that for this particular DIB
carrier, rotational excitation is not very efficient.

\acknowledgements We would like to thank Bernard Foing and Xander
Tielens for stimulating discussions, and the anonymous referee for
making valuable suggestions.  J.~C. gratefully acknowledges the
support of the National Research Council's Research Associateship
Program. This work is supported by the NASA APRA Program (RTOP
188-01-03-01). J.~K. is grateful to the Polish State Committee for
Scientific Research for support under grant 2 P03D 019 23. G.~G is
grateful to KOSFT for providing an opportunity to work at the Korea
Astronomy Observatory through the Brain Pool program and acknowledges
the Ministry of Science and Technology, Korea, for support under
grant M1-022-00-0005 and the Russian Foundation for Basic Research
for financial support under grant 02-02-17423.


\end{document}